\documentclass[useAMS,usenatbib]{mn2e}

\usepackage{amssymb}
\usepackage{amsfonts}
\usepackage{mncite}
\usepackage{epsfig}
\usepackage{psfig}

\begin{document}

\title[Tidal evolution of close-in planets]
{Tidal evolution of close-in giant planets : Evidence of type II migration?}

\author[W.K.M. Rice, J. Veljanoski, \& A. Collier Cameron]
 {W.K.M. Rice$^1$\thanks{E-mail: wkmr@roe.ac.uk}, J. Veljanoski$^1$, 
A. Collier Cameron$^2$ \\
$^1$ SUPA\thanks{Scottish Universities Physics Alliance},
Institute for Astronomy, University of Edinburgh, Blackford Hill, Edinburgh, EH9 3HJ \\
$^2$ SUPA, School of Physics and Astronomy, University of St Andrews, North Haugh, St Andrews, Fife KY169SS}

\maketitle

\begin{abstract}
It is well accepted that `hot Jupiters' and other short-period planets did not form {\em in situ}, as the temperature in the protoplanetary disc
at the radius at which they now orbit would have been too high for planet formation to have occurred.  These planets, instead, form
at larger radii and then move into the region in which they now orbit.  The exact process that leads to the formation
of these close-in planets is, however, unclear and it seems that there may be more than one mechanism that can produce 
these short-period systems. Dynamical interactions in multiple-planet systems can scatter planets into highly eccentric
orbits which, if the pericentre is sufficiently close to the parent star, can be tidally circularised by tidal interactions
between the planet and star.  Furthermore, systems with distant planetary or stellar companions can undergo Kozai cycles
which can result in a planet orbiting very close to its parent star. However, the most developed model for the origin of short period
planets is one in which the planet exchanges angular momentum with the surrounding protoplanetary disc and spirals in towards
the central star.  In the case of `hot Jupiters', the planet is expected to open a gap in the disc and migrate in, what is known as, the Type II regime.
If this is the dominant mechanism for producing `hot Jupiters' then we would expect the currect properties of observed close-in 
giant planets to be consistent with an initial population resulting from Type II migration followed by evolution due to
tidal interactions with the central star.  We consider initial distributions that are consistent with Type II migration and find
that after tidal evolution, the final distributions can be consistent with that observed.  Our results suggest that a modest initial pile-up at $a \sim 0.05$ au
is required and that the initial eccentricity distribution must peak at $e \sim 0$. We also suggest that if higher-mass close-in exoplanets
preferentially have higher eccentricities than lower-mass exoplanets, this difference is primordial and is not due to subsequent
evolution. 
\end{abstract}

\begin{keywords}
stars: formation --- stars: pre-main-sequence --- circumstellar matter --- planetary systems: protoplanetary discs ---
planetary systems: formation
\end{keywords}

\section{Introduction}
The first extrasolar planet detected around a solar-like star was 51 Pegasi b, discovered in 1995 \citep{mayor95}. This planet, with a mass
about half that of Jupiter and a semimajor axis of $0.052$ au, was the prototype of a class of planets now known
as `hot Jupiters'.  These are gas giant planets that orbit close ($a \le 0.1$ au) to their parent stars.  
There is, however, a general consensus that these planets did not form {\it in situ} since the temperature in the protoplanetary disc at the
radii where these now orbit would be too high for planet formation to proceed \citep{bell97}.  

The most developed model
for the origin of `hot Jupiters' and other short-period planets is one in which these planets exchange angular momentum
with the surrounding protoplanetary disc and spiral in towards the central star \citep{goldreich80,lin86}.  In the case of
`hot Jupiters' the planet is expected to open a gap in the disc and migrate in what is known as the Type II regime.  Lower-mass,
close-in planets may migrate in the gapless Type I regime \citep{ward97}. This work will, however, only be considering
planets that could potentially have undergone Type II migration.

Migration is not the only mechanism that can lead to planets orbiting very close to their parent stars.  Dynamical interactions 
in multiple planet systems - often referred to as planet-planet scattering - can scatter planets into highly eccentric orbits which, if the pericentre is sufficiently close
to the star, can be circularised by tidal interactions between the planet and the host star \citep{rasio96}. 
Furthermore, systems with distant stellar or planetary companions on inclined orbits (with respect to the inner planet's
orbit) can undergo Kozai cycles which, if then followed by 
tidal evolution, can result in a planet orbiting very close to its parent star \citep{kozai62, eggleton98, wu03}.   
 
These different mechanism will result in different distributions of close-in planets. Early simulations 
of planet-planet scattering \citep{ford01} found that the likelihood of a massive planet being scattered into an orbit that
brought it very close to the central star was less than $1$ \%, suggesting that this was unlikely to be 
the primary mechanism for producing `hot Jupiters'. Longer term integration \citep{marzari02} increases this
to $\sim 10$ \%, while \citet{nagasawa08} suggest that ``Kozai migration" \citep{wu03} can further increase this
to $\sim 30$ \%. Recent observations \citep{winn09,colliercameron10, queloz10} of exoplanets with orbits
that are retrograde with respect to the rotation of the their central star suggests that ``Kozai migration"
must play a role in the formation of some close-in planets \citep{triaud10}.  

Theoretical modelling \citep{fabrycky07} and analysis of the current sample of close-in exoplanets \citep{morton11}
suggests that Kozai cycles together with tidal friction can produce some, but not most, of the close-in planets. In this work 
we intend to investigate if initial conditions consistent with Type II migration of giant planets can lead - after tidal
evolution - to distributions that are consistent with those observed.  Although it is fairly clear that Type II migration
is not the only mechanism capable of producing close-in giant planets, if it is the dominant mechanism we would at least expect
the current distribution to be consistent with one having evolved from an initial distribution resulting from Type II migration. 
The paper is structured as follow.  Section 2 describes the basic model and assumptions, Section 3 dicusses the results
and in Section 4 we discuss the results and draw conclusions.

\section{Basic Model}
In this paper we use knowledge of Type II migration to choose initial semi-major and eccentricity distributions
of gas giant planets around Sun-like stars and evolve these systems by integrationg the equations that describe
the tidal evolution of these systems.

\subsection{Tidal evolution}
We base our model of tidal evolution on that described in detail in \citet{dobbs04}.
We reproduce the relevant parts here. If the spin axes of a star and planet are aligned, tidal energy will be dissipated
at rates defined by the tidal quality factors $Q'_*$ and $Q'_p$. 
Consider a planet of mass $M_p$ and radius $R_p$ orbiting
a star of mass $M_*$ and
radius $R_*$.  If the system has a semimajor axis $a$ and angular 
frequency $n = \left(G (M_*+M_p)/a^3 \right)^{1/2}$, the rate of change
of the eccentricty of the orbit is given by \citep{eggleton98,mardling02,dobbs04}
\begin{equation}
\dot{e} = g_p + g_*,
\label{edot}
\end{equation}
where
\begin{eqnarray}
g_{p,*} = \left(\frac{81}{2} \frac{n e}{Q'_{p,*}} \right) \left(\frac{M_{*,p}}{M_{p,*}} \right)
\left(\frac{R_{p,*}}{a}\right)^5 \times \nonumber \\
\left[-f_1(e) + \frac{11}{18} f_2(e) 
\left( \frac{\Omega_{p,*}}{n} \right) \right],
\label{gp*}
\end{eqnarray}
\begin{equation}
f_1(e) = \left(1 + \frac{15}{4}e^2 + \frac{15}{8}e^4+\frac{5}{64}e^6\right)/\left(1-e^2 \right)^{13/2},
\label{f1}
\end{equation}
\begin{equation}
f_2(e)=\left(1 + \frac{3}{2}e^2+\frac{1}{8}e^4\right)/\left(1-e^2\right)^5.
\label{f2}
\end{equation}

The stellar and planetary spin can also play an important role in the tidal evolution
of close in planets.  If the rotation axes are aligned and if $M_* >> M_p$, the
rate of change of the stellar spin frequency is \citep{mardling02,dobbs04}
\begin{eqnarray}
\dot{\Omega_*} = \frac{9}{2} \left( \frac{n^2}{\epsilon_* \alpha_* Q'_*} \right) 
\left(\frac{M_p}{M_*}\right)^2\left(\frac{R_*}{a}\right)^3 \times \nonumber \\
\left[f_3(e) - f_4(e)\left(\frac{\Omega_*}{n}\right)\right] + \dot{\omega_*},
\label{omegastardot}
\end{eqnarray}
where
\begin{equation}
f_3(e) = \left(1 + \frac{15}{2}e^2 + \frac{45}{8}e^4 + \frac{5}{16}e^6\right)/(1-e^2)^6,
\label{f3}
\end{equation}
\begin{equation}
f_4(e) = \left( 1 + 3 e^2 + \frac{3}{8} e^4 \right)/(1 - e^2)^{9/2},
\label{f4}
\end{equation}
and $\dot{\omega_*}$ is the change of stellar rotation due to angular momentum loss through
a stellar wind, which we will discuss in more detail in a later section. The quantity
$\epsilon_*$ is the fraction of the mass of the star participating in tidal exchange,
and $\alpha_*$ is the star's moment of inertia coefficient.  We use $\epsilon_* = 0.1$, 
appropriate for G stars which have shallow convection zones, and $\alpha_* = 0.1$. 

The rate of change of the planet's spin is 
\begin{equation}
\dot{\Omega_p} = \frac{9}{2} \left( \frac{n^2}{\epsilon_p \alpha_p Q'_p} \right)
\left(\frac{M_*}{M_p} \right) \left( \frac{R_p}{a} \right)^3
\left[ f_3(e) - f_4(e) \left(\frac{\Omega_p}{n} \right) \right].
\label{omegaplanetdot}
\end{equation}
The quantities $\epsilon_p$ and $\alpha_p$ are the fraction of the planet's mass involved
in tidal exchange, and the planet's moment of inertia coefficient. Since gas giant planets 
can be fully convective, we use $\epsilon_p  = 1$ and use $\alpha_p$ = 0.2. 

The total angular momentum perpendicular to the orbit is
\begin{equation}
J_{\rm total} = J_o(a,e) + J_p(\Omega_p) + J_*(\Omega_*).
\label{totalangmom}
\end{equation}
The orbital angular momentum is $J_o(a,e) = M_p M_* a^2 n (1-e^2)^{1/2}/(M_p + M_*)$,
while the angular momentum of the star and planet are
\begin{eqnarray}
J_* &=& \alpha_s \epsilon_* M_* R_*^2 \Omega_*, \nonumber \\
J_p &=& \alpha_p M_p R_p^2 \Omega_p,
\label{angmomstarplanet}
\end{eqnarray}
where it is assumed that only a fraction $\epsilon_*$ of the star, but all of the planet, is involed 
in angular momentum exchange. The rate of change
of the total angular momentum is then equal to the rate at which the system loses angular
momentum through the stellar wind, $\dot{J_{\omega_*}}$. Differentiating equation
(\ref{totalangmom}) and using $\dot{J_{\omega_*}} = \alpha_* \epsilon_* M_* R^2_* \dot{\omega_*}$
gives
\begin{eqnarray}
\alpha_* \epsilon_* M_* R^2_* \dot{\omega_*} = J_o \left( \frac{\dot{a}}{2 a}
- \frac{e \dot{e}}{1 - e^2} \right) + \nonumber \\
\alpha_p M_p R^2_p \dot{\Omega_p} +
 + \alpha_* \epsilon_* M_* R^2_* \dot{\Omega_*}
\label{angmomrate}
\end{eqnarray} 
The rate of change of the semi-major axis of the orbit can then be determined by simply
rearranging equation (\ref{angmomrate}).

\subsection{Stellar wind model}
We take the stellar wind model from \citet{colliercameron94}. In the
unsaturated dynamo regime with purely thermal driving, the rate
of change of a star's angular frequency is
\begin{equation}
\dot{\omega_*} = - \kappa \Omega_*^3,
\label{wind}
\end{equation} 
where
\begin{equation}
\kappa = \frac{2}{3} B^2_{\rm 0, \odot} \left( \frac{\tau_c}{\tau_{c, \odot} \Omega_\odot} \right)^2
\left( \frac{\beta m_{\rm p}}{2 k_{\rm B} T_{\rm w}} \right)^{1/2} \frac{R_*^2}{k^2 M_*}.
\label{kappa}
\end{equation}
For the solar values we assume a magnetic flux density of $B_{\rm 0, \odot} = 3$ G,
a convective turnover time of $\tau_{c, \odot} = 8.9 \times 10^5$ s, and an
angular velocity of $\Omega_{\rm 0, \odot} = 4.0 \times 10^6$ rad s$^{-1}$. To determine 
$\tau_c$ we assume a linear fit to the values in Table 1 of \citet{colliercameron94}.
The quantities $m_{\rm p}$ and $k_{\rm B}$ are the proton mass and Boltzmann's constant
and from \citet{mestel87} we assume $\beta = 0.16$.  The final quantity $k$ is the
effective radius of gyration of the radiative interior and convective envelope
combined, which we take to be $k^2 = 0.1$ \citep{colliercameron94}.

In the case of stars with saturated dynamos,
\begin{equation}
\dot{\omega_*} = - \kappa \tilde{\Omega}^2 \Omega_*,
\label{satwind}
\end{equation}
where $\tilde{\Omega}$ is the saturation limit which we get by assuming  linear fit to
the values in Table 5 of \citet{colliercameron94}.

\subsection{Basic setup}
We carry out Monte Carlo simulations of a large sample of close-in planets ($a < 0.3$ au) around
solar-like stars. We consider a
range of different initial radial distributions and eccentricity distributions that we
will discuss in more detail in a later section.  There are, however, some common elements to
all of the simulations that we discuss here.

Based on the properties of the known exoplanets we distribute the planet masses as $M_p \propto
M^{-1.15}$ \citep{marcy08} with a minimum planet mass of $M_p = 0.1$ M$_{\rm Jup}$ and 
a maximum planet mass of $M_p = 15$ M$_{\rm Jup}$. The minimum and maximum masses are chosen such as to represent  
the range of planet masses that could undergo Type II migration \citep{d'angelo03}.
The planet radii are computed using $R \propto M^{\frac{1-m}{3-m}}$ 
where $m$ is varied from $m=0$ for low-mass planets to $m = 1.5$ for the most massive gas giants,
to give a mass-radius relation that matches that expected theoretically \citep{chabrier09} and
that is normalised to give $R_p = 1$ R$_{\rm Jupiter}$ when $M_p = 1$ M$_{\rm Jupiter}$.
The stars are assumed to have masses evenly distributed between 0.75 M$_\odot$ and 1.25 M$_\odot$ and
have rotation periods distributed as a Gaussian 
with a peak at 8 days and with a half-width of $1.5$ days. 
The planets are all assumed to start with rotation rates 10 times greater
than their orbital periods (i.e., $\Omega_p = 10 n$). 

In each simulation we consider 1800 planets. The initial semi-major axis, eccentricity, planet
mass, stellar mass, and stellar rotation period are all chosen randomly, but in such a way
that the final distributions match those described above (we will discuss the chosen
semi-major axis and eccentricity distributions below). The star and planet radii
are then determined as above and the star mass and rotation periods 
determine - using equations (\ref{wind}) and
(\ref{kappa}) -  the rate at which the star 
loses angular momentum through a stellar wind.  For the star and planet, we use the tidal 
quality factors of $Q'_s = 10^6$ and $Q'_p = 5 \times 10^6$.  These are similar to those used 
by \citet{jackson09} and \citet{dobbs04} and consistent with those determined by \citet{brown11}.  
We did try using the tidal quality factors estimated by \citet{hansen10},
but these resulted in circularisation timescale for planets with $4$ and $5$ day periods that were much 
longer than expected.

Equations (\ref{edot}), (\ref{omegastardot}), (\ref{omegaplanetdot}),
and (\ref{angmomrate}) - with equation (\ref{angmomrate}) rewritten in terms of $\dot{a}$ - are
then integrated using a fourth-order Runge Kutta integrator for a time chosen randomly between
$2 \times 10^9$ years and $6 \times 10^9$ years. Any planet that reaches its Roche
limit, given by $a_R = (R_p/0.462)(M_*/M_p)^{1/3}$ \citep{faber05}, is assumed to be
tidally disrupted and destroyed \citep{jackson09}. 

\section{Results}
We assume here that close-in giant planets are unable to form in situ, but rather form beyond
the snowline ($a \ge 2.7$ au for a Solar-mass star) and migrate inwards to their current locations \citep{lin96,trilling98}. Although there are
a number of mechanism that could result in planet migration, in the case of giant planets 
the most likely mechanism is Type II migration in which the planet migrates within a
gap in the proplanetary disc \citep{goldreich80,lin86}. Using theoretical models of 
Type II migration \citet{armitage07} (following on from \citet{armitage02}) has shown
that the resulting radial distribution of extrasolar planets is consistent with that 
observed, the best fit occuring if migration is assumed to be slightly suppressed  
at small radii. Although this ignores some effects, in particular planet-planet scattering which
can also change the orbital elements of surviving planets \citep{ford01}, it is
at least a reasonable starting point.

\begin{figure}
\begin{center}
\psfig{figure=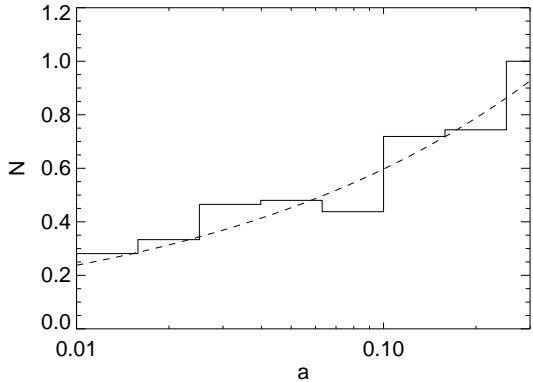,width=0.45\textwidth} 
\end{center}
\caption{Initial radial distribution determined by assuming Type II inward
migration with migration slightly suppressed at small radii \citep{armitage07}.
Also shown is the curve $d N / d \log a \propto a^{0.4}$ for comparison. }
\label{a_initial}
\end{figure}

\begin{figure}
\begin{center}
\psfig{figure=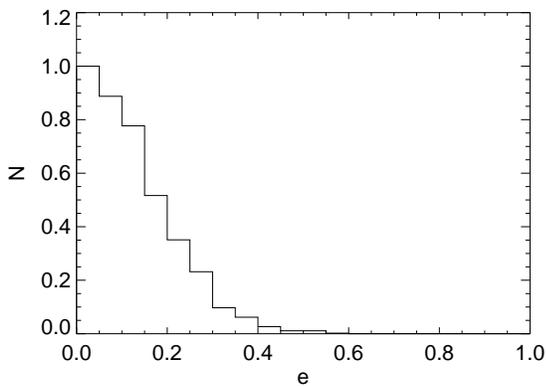,width=0.45\textwidth}
\end{center}
\caption{Initial eccentricity distribution used in the simulations presented. The final
radial distributions do not, however, depend particularly strongly on the initial eccentricity 
distribution.}
\label{e_initial}
\end{figure}

Inside $1$ au, the best-fit model of \citet{armitage07} has a radial distrbution of 
$dN/d \log a \propto a^{0.4}$. We therefore use this, shown in Figure \ref{a_initial},
as the base distribution for all of our simulations. We consider a few different
initial eccentricity distributions, but since the final radial distributions
don't depend strongly on the initial eccentricity distribution \citep{jackson09},
most of our simulations use the intial eccentricity distribution shown in Figure \ref{e_initial}. 

\subsection{Type II migration only}
\begin{figure}
\begin{center}
\psfig{figure=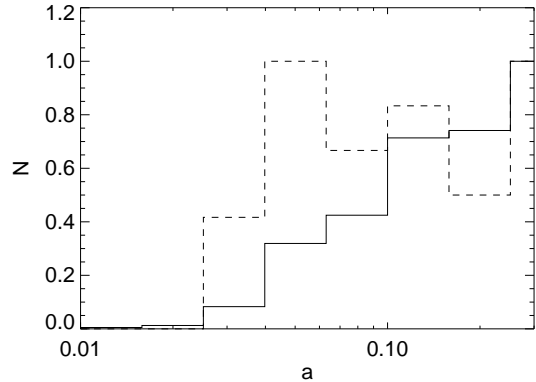,width=0.45\textwidth}
\end{center}
\caption{The final semi-major axis distribution for the simulated planets 
(solid line) compared to the semi-major axis distributed for observed
exoplanets (dashed line) in systems with properties comparable to those
considered in the simulations.}
\label{a_final}
\end{figure}

Our first set of simulations consider the case of Type II migration with slight 
suppression at small radii \citep{armitage07}.
We therefore use the initial distribution shown in Figure \ref{a_initial} and
integrate each system for a randomly chosen time of between $2$ and $6$ Gyr.  The
final radial distribution is shown in Figure \ref{a_final}. This illustrates 
that the tidal interaction causes most of the planets inside $0.04$ au
to migrate inwards ultimately overflowing their Roche lobes and being destroyed
\citep{jackson09}.
Outside $0.04$ au, however, the distribution is not significantly different to the initial
distribution.  

\begin{figure}
\begin{center}
\psfig{figure=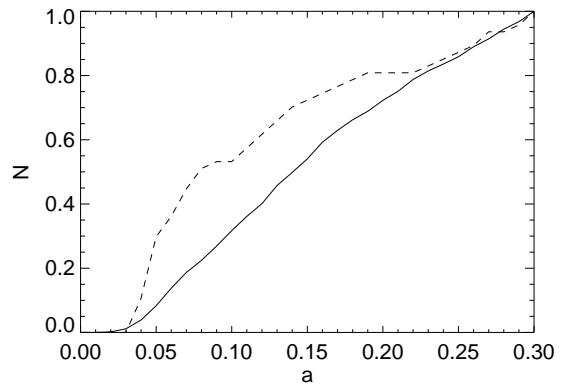,width=0.45\textwidth}
\end{center}
\caption{Cumulative semi-major axis distribution for the simulated exoplanets (solid line) and
for the observed exoplanets (dashed line), showing that the radial distribution of the observed
exoplanets does not appear to be consistent with the simulated systems that are taken to have an initial
radial distribution resulting from Type II migration alone.}
\label{cumul_distr}
\end{figure}

The dashed-line in Figure \ref{a_final} shows the radial distribution of known
close-in exoplanets. Since we are only interested in those that are likely to undergo
Type II migration, we only include exoplanets for which $M \ge 0.1$ Jupiter masses.
We have included only those
planets discovered using the doppler radial velocity technique \citep{mayor95,udry07}. 
We considered also including planets detected via transits, but were concerned that selection
effects may somewhat enhance any pile-up at small radii.  So that our simulated 
sample and the observed sample can be compared, we have also excluded any planets in our
simulated sample that would induce a stellar radial velocity of less than $2$ m s$^{-1}$,
and only included those observed exoplanets whose host star masses are between
$0.75$ and $1.25$ M$_\odot$.  Figure \ref{a_final} suggests that these two distributions
are not the same, with some evidence for a pile-up of observed exoplanets at 
around $0.05$ au, that is not seen in the simulated distribution.  Figure \ref{cumul_distr}
shows the cumulative distribution for the two samples and again illustrates that a
larger fraction of the observed exoplanets are located inside $0.1$ au, 
compared to the simulated systems. Formally a Kolmogorov-Smirnoff test shows that
the probability that the two distributions are drawn from the same parent
distribution is $P_{\rm KS} = 0.36$. 

\subsection{Type II migration with stopping mechanism}
It has been suggested that rather than migrating all the way in to
the central star, planets may be prevented from doing so by some kind
of stopping mechanism that halts planets with periods of $\sim 4$ days,
corresponding to a semimajor axis of $\sim 0.05$ au. One possibility 
is that the magnetic fields of TTauri stars may be
strong enough to disrupt the inner edge of protoplanetary
discs \citep{konigl91,bouvier07}, and will truncate the disc
close to the corotation radius. Planets will continue to migrate
inside this cavity, but this is expected to slow and become 
very inefficient once planets
pass inside the 2:1 resonance with the inner edge of the
disc \citep{lin96, kuchner02, rice08}. The distribution of stellar
rotation periods \citep{herbst07}, with a peak at 
$\sim 8$ days, would therefore produce a pile-up
of planets with orbital periods $\sim 4$ days, corresponding 
to an orbital radius of $\sim 0.05$ au around a solar-like star.

\begin{figure}
\begin{center}
\psfig{figure=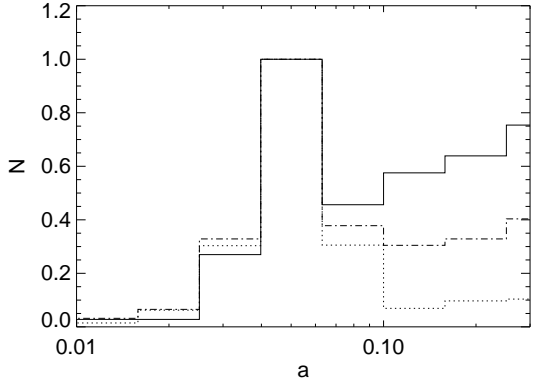,width=0.45\textwidth}
\end{center}
\caption{The initial radial distribution of planets assuming Type II inward
migration followed by a pile-up of planets inside $0.1$ au with the
peak of the pile-up occuring where the orbital
period is half the stellar spin period ($a \sim 0.05$ au), and a width corresponding to
a variation in the orbital period of $\pm 2$ days. The three cases shown are 
one in which the number of planets inside $0.1$ is the same as in Figure \ref{a_initial} (solid line),
an enhancement of $20$ \% over that in Figure \ref{a_initial} (dash-dot line), and a further enhancement
of $30$ \% (dotted line).}
\label{a_initial_peaked}
\end{figure}

As discussed above, \citet{armitage07} suggests that Type II
migration would lead to a radial distribution, within $1$ au, approximated by 
$dN/d \log a \propto a^{0.4}$.  This distribution would suggest
that $\sim 55$ \% of the planets between $r = 0.01$ and $r = 0.3$ au
would be located insided $0.1$ au. We therefore consider 3 different
pile-up scenarios, illustrated in Figure \ref{a_initial_peaked}.  
The solid line is a case in which the number of planets between $r = 0.01$ and $r = 0.1$ au
is the same as would be expected from the radial distribution derived
by \citet{armitage07}.  The dash-dot line increases the fraction to $65$ \%, 
and the dash-dot-dot line is a case in which the 
fraction inside $0.1$ au is enhanced to $\sim 88$ \%.   

\begin{figure}
\begin{center}
\psfig{figure=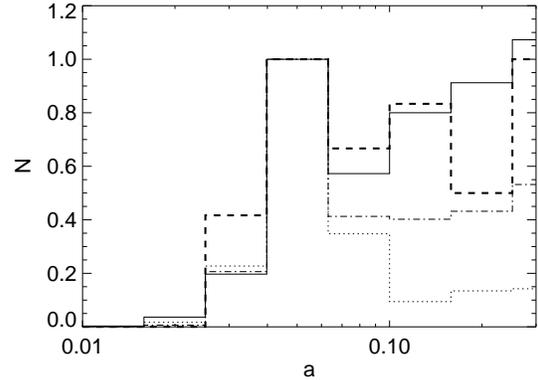,width=0.45\textwidth}
\end{center}
\caption{The final radial distribution of the simulated systems 
with initial radial distributions as shown in Figure \ref{a_initial_peaked} 
(solid, dash-dot and dotted lines ) compared to the observed radial distribution of exoplanets (thick dashed line).}
\label{a_final_peaked}
\end{figure}

\begin{figure}
\begin{center}
\psfig{figure=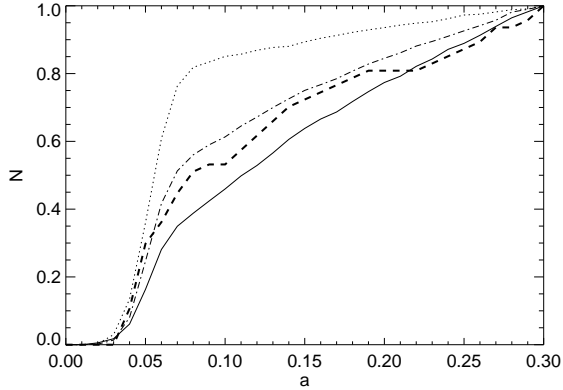,width=0.45\textwidth}
\end{center}
\caption{Cumulative semi-major axis distribution for the simulated systems (solid, dash-dot and dotted lines) with
with initial radial distributions shown in Figure \ref{a_initial_peaked}, compared to the
cumulative radial distribution of the observed exoplanets (thick dashed line). This appears
to illustrate that the observed distribution is consistent with a primordial pile-up at $r \sim 0.05$ au but
not if there is a significant enhancement in the fraction of planets inside $0.1$ au.}
\label{cumul_distr_peaked}
\end{figure}

As before, we evolve each system for a randomly chosen time of between $2$ and $6$ Gyr. The
final radial distributions are shown in Figure \ref{a_final_peaked}. The thick dashed-line is again
the radial distribution of the known exoplanets and again we have only included those 
discovered using the doppler radial velocity technique and that orbit stars with masses between $0.75$ and $1.25$
M$_\odot$. We have also removed any simulated planet that would induce a radial velocity of
less than $2$ m s$^{-1}$.  The line styles in Figure \ref{a_final_peaked} correspond to those shown in 
Figure \ref{a_initial_peaked}.  In all three case, the pile up remains but the best fit with the observed 
distribution is somewhere between the solid and dash-dot line.  The cumulative distributions are shown
in Figure \ref{cumul_distr_peaked}.  The line styles again correspond to those in Figure \ref{a_initial_peaked}
and Figure \ref{a_final_peaked}.  Again this illustrates that the best fit to the observed distribution would be 
somewhere between the solid line and the dash-dot line.  
The Kolmogorov-Smirnoff test shows that
both the solid and dash-dot lines, when compared with the dashed line, 
have a probability of $P_{\rm KS} = 0.94$ of being 
drawn from the same parent distribution.  

The above result suggests that, to explain the current distribution of close-in giant planets, there must have
been a primordial pile-up of giant planets with the peak of the pile-up occuring at $\sim 0.05$ au. The best
fit also occurs if the number of planets between $0.01$ and $0.1$ is similar to, but slightly higher than, that expected based on
the radial distribution of planets resulting from Type II migration \citep{armitage07}.  One has to be slightly
careful with this comparison, however, as the radial profile determined by \citet{armitage07} is a steady-state
profile and does not indicate what fraction of planets will have migrated inside $0.1$ au and then been lost.   
Population synthesis models \citep{ida04} suggest that a large fraction of giant planets
should migrate close to their host stars.  If the stopping mechanism
were efficient, we would then expect a significant primordial pile-up at small
radii.  Figures \ref{a_final_peaked} and \ref{cumul_distr_peaked} are consistent with
a pile-up but are not consistent with a particularly significant pile-up.
This suggests either that the
stopping mechanism is not efficient and that planets are lost before tidal evolution becomes important, 
or that a smaller fraction of giant planets migrate close to their parent
star than predicited by population synthesis models.

\subsection{Mass and eccentricity dependence}
\begin{figure}
\begin{center}
\psfig{figure=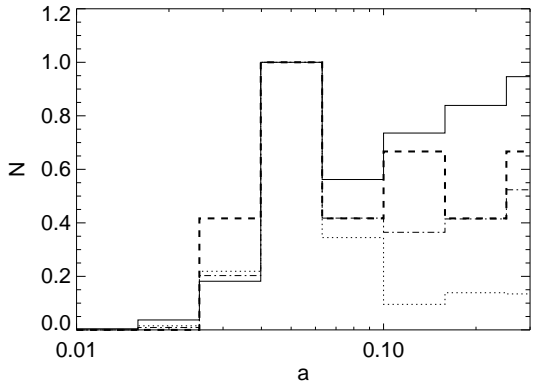,width=0.45\textwidth}
\end{center}
\caption{The final radial distribution of the simulated systems
with initial radial distributions as shown in Figure \ref{a_initial_peaked}
(solid, dash-dot and dotted lines ) and with masses $M_p < 2$ M$_{\rm Jup}$, compared to the observed radial 
distribution of exoplanets with the same range of masses (thick dashed line).}
\label{a_final_peaked_lowmass}
\end{figure}

\begin{figure}

\begin{center}
\psfig{figure=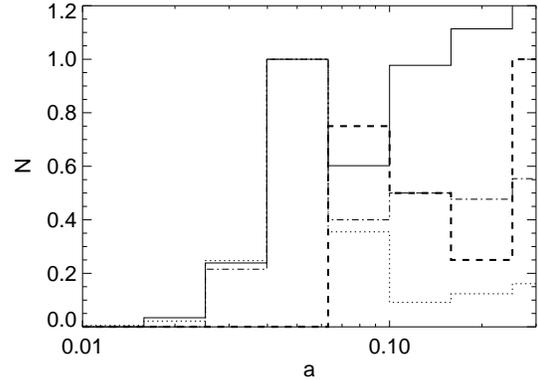,width=0.45\textwidth}
\end{center}
\caption{The final radial distribution of the simulated systems
with an initial radial distributions as shown in Figure \ref{a_initial_peaked}
(solid, dash-dot and dotted lines ) and with masses $M_p > 2$ M$_{\rm Jup}$, 
compared to the observed radial distribution of exoplanets with the same range of masses (thick dashed line).
The simulated systems still show a peak at $\sim 0.05$ au that isn't present in the observed systems.}
\label{a_final_peaked_highmass}
\end{figure}

In Figure \ref{a_final_peaked} we consider the entire mass range from $M_p = 0.1$ M$_{\rm Jup}$ to a maximum of
$M_p = 15$ M$_{\rm Jup}$.  If, however, we divide this into low-mass regime ($M_p < 2$ M$_{\rm Jup}$) and a high-mass
regime ($M_p > 2$ M$_{\rm Jup}$) we find (Figure \ref{a_final_peaked_lowmass}) that for the low-mass planets the
peak at $a \sim 0.05$ au remains and is still well matched for our simulated populations that have modest pile-ups
at this radius.  For the high-mass planets (Figure \ref{a_final_peaked_highmass}) this is no longer the case.  In the
observed population there are no planets with $a \sim 0.05$ au.


It has been suggested \citep{rice08} that the interaction between a planet, located inside
a magnetospheric cavity, and the surrounding disc can lead to substantial eccentricity
growth for planets with $M > 1$ M$_{\rm Jup}$ and that the growth rate is 
greatest for the highest mass planets.   It was therefore suggested \citep{rice08} that this would 
result in higher-mass, close-in planets being preferentially destroyed when compared with lower-mass,
close-in planets.  We therefore consider a situation
in which the initial radial distribution is as shown in Figure \ref{a_initial_peaked}, but the
initial eccentricities increase with increasing mass.  We assume, somewhat arbitrarily, the planets with
masses below 1 $M_{\rm Jup}$ have an eccentricity peak close to $e = 0$, but have a tail that can extend 
to $e = 0.6$, while planets with masses above $10 M_{\rm Jup}$ have eccentricities that lie preferentiall
between $e=0.4$ and $e = 0.8$.
We assume a maximum eccentricity of $e = 0.8$ so that we don't simply lose a large
fraction of the massive exoplanets immediately.  We should note that simulations by \citet{benitez11} did not find
eccentricity growth for massive planets in inner disc gaps.  Their simulations, however, did not have completely
evacuated inner cavities and so differ from those of \citet{rice08}. 

\begin{figure}
\begin{center}
\psfig{figure=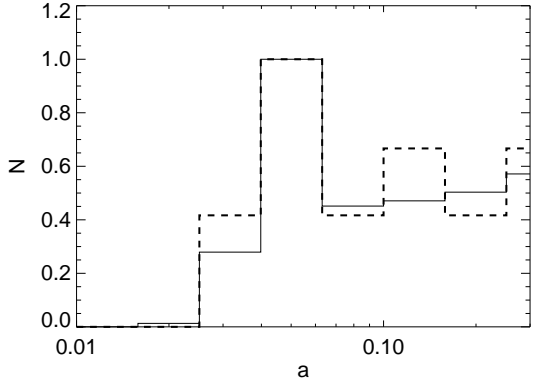,width=0.45\textwidth}
\end{center}
\caption{The final radial distribution of the simulated planets with masses $M_p < 2$ M$_{\rm Jup}$ and 
with an initial peaked radial distribution illustrated by the solid line in Figure \ref{a_initial_peaked}.
The initial eccentricity distribution is also assumed to depend on planet mass with the higher-mass planets preferentially
having higher eccentricities than the lower-mass planets.  For these lower- mass planets the fit to the observed distribution 
(thick dashed line) is still good.}
\label{a_final_peaked_massdep_lowmass}
\end{figure}

\begin{figure}

\begin{center}
\psfig{figure=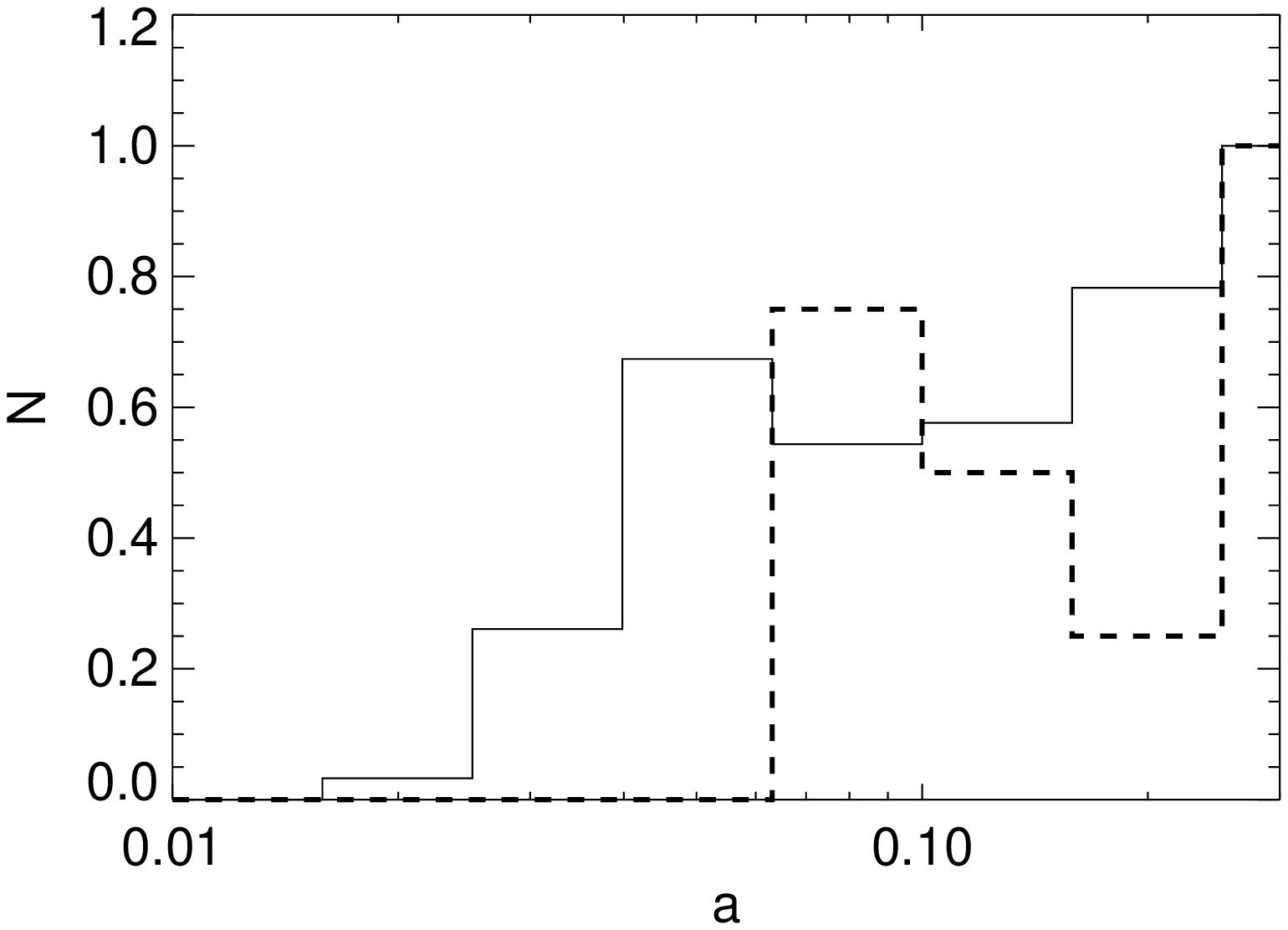,width=0.45\textwidth}
\end{center}
\caption{The final radial distribution of the simulated planets with mass $M_p > 2$ M$_{\rm Jup}$ and 
with an initial peaked radial distributions illustrated by the solid line in Figure \ref{a_initial_peaked}.
The initial eccentricity distribution is also assumed to depend on planet mass with the higher-mass planets preferentially
having higher eccentricities than the lower-mass planets.  Although, for these higher-mass planets, the fit to the observed distribution
(thick dashed line)
is not particularly good, it is clear that the higher initial eccentricities has largely removed the initial peak at $a \sim 0.05$ au.}
\label{a_final_peaked_massdep_highmass}
\end{figure}

\begin{figure}
\begin{center}
\psfig{figure=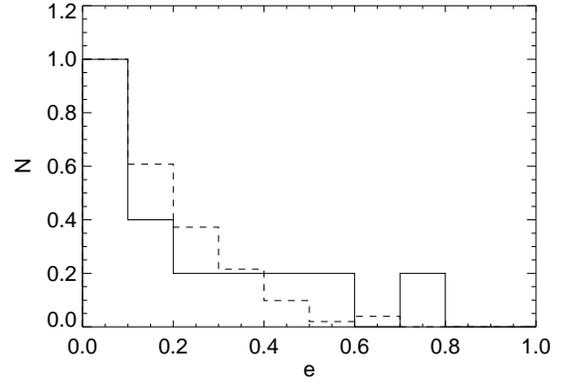,width=0.45\textwidth}
\end{center}
\caption{Eccentricity distribution for high-mass ($M > 2 M_{\rm Jup}$, solid line) and
low-mass ($M < 2 M_{\rm Jup}$, dashed line) exoplanets, detected via radial velocity, with semimajor axes less than
$0.3$ au. There is a suggestion of
a group of higher-mass exoplanets that retain relatively large eccentricities.}
\label{e_final_obsonly}
\end{figure}

\begin{figure}
\begin{center}
\psfig{figure=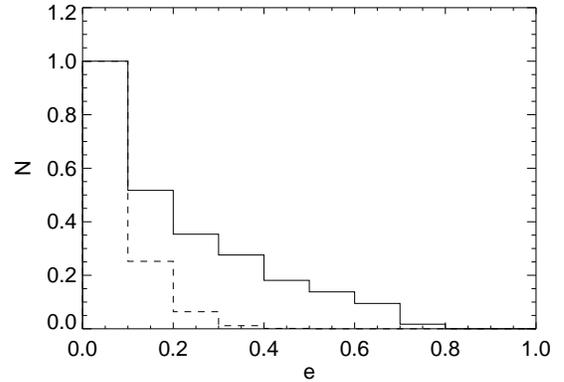,width=0.45\textwidth}
\end{center}
\caption{Simulated final eccentricity distribution for low-mass
($M < 2 M_{\rm Jup}$, dashed line) and high-mass ($M > 2 M_{\rm Jup}$, solid line) planets, with
an initial mass dependent eccentricity distribution. In this case, a group of
high-mass planets with large eccentricities remains and is at least qualitatively
consistent with that observed.}
\label{e_final_massdep}
\end{figure}

Figure \ref{a_final_peaked_massdep_lowmass} shows the resulting radial distribution for the lower-mass planets 
($M_p < 2$ M$_{\rm Jup}$ - solid line) which still agrees well with the observed radial distribution (thick dashed line).  
The radial distribution for the higher-mass planets ($M_p > 2$ M$_{\rm Jup}$ - solid line) is shown in 
Figure \ref{a_final_peaked_massdep_highmass}, and although not a particularly good fit
to the observed distribution (thick dashed line) does at least indicate that the pile-up is significantly depleted. 
Furthermore, if one considers the final eccentricity distribution, shown in Figure \ref{e_final_massdep}, again divided into planets
with masses above $2$ M$_{\rm Jupiter}$ (solid line) and planets with masses below $2$ M$_{\rm Jupiter}$
(dashed line),  a group of high-mass exoplanets with large eccentricities remains and
is at least qualitatively similar to that observed (Figure \ref{e_final_obsonly}).  This suggests that if higher-mass exoplanets do indeed
preferentially have higher eccentricities than lower-mass exoplanets, this could explain why the pile-up at $0.05$ au is only
observed for the lower-mass exoplanets and  suggests that the mass dependent eccentricity distribution must be primordial
rather than being due to a difference in their subsequent evolution \citep{rice08}.


\subsection{Planet-planet scattering}
The main goal of this work has been to show that an initial distribution consistent with
that expected from Type II migration \citep{armitage07} followed by a pile-up due to truncation
of the inner disc \citep{rice08} is at least consistent - after tidal evolution - with the
observed distribution of close-in exoplanets.  We cannot, however, preclude the possibility
that migration occurs through other mechanisms. 

\citet{wright09} have shown that there
is a significant difference in the semi-major axis distribution of single
and multiple-planet systems. The single-planet systems have a pile-up at $\sim 3$ days
and a radial distribution that increases with increasing radius.  The multiple-planet systems,
on the other hand, have a semi-major axis distribution that is only weakly dependent on radius.
It has been suggested (e.g., \citet{matsumura10}) that this difference could be
due to single-planet systems being dominated by planet-planet scattering (with
additional planets being ejected from the system) while multiple-planet systems
are dominated by migration in a gas disc.  Type II migration would, according to
\citet{armitage07}, result in 
a semi-major axis distribution that increases 
significantly with increasing radius.  Qualitatively,
this is more consistent with the radial distribution of the single-planet systems than it is
with the multiple-planet systems, which have a semi-major axis distribution that is only 
weakly dependent on radius.

\begin{figure}
\begin{center}
\psfig{figure=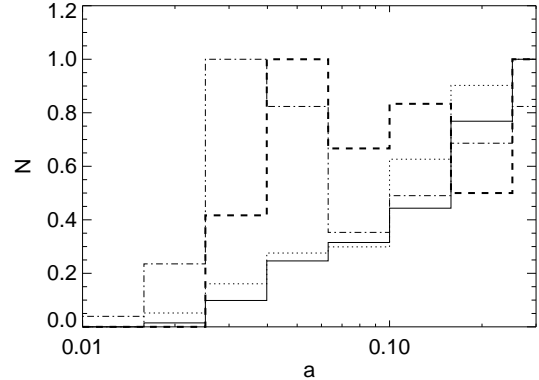,width=0.45\textwidth}
\end{center}
\caption{The final radial distribution of the simulated systems
with initial eccentricity distributions that peak at $e \sim 0.6$ and with initial radial distributions
that are unpeaked (solid line), weakly peaked (dotted line) and very strongly peaked (dash-dot line). 
Also shown is the observed radial distribution of exoplanets (thick dashed line). A peak only remains for the simulated
system which was initially strongly peaked (dash-dot line) suggesting that, if the pile-up at $\sim 3$ days is primarily due to
planet-planet scattering, it must be very effective at scattering planets into orbits with $a < 0.1$ au.}
\label{a_final_highe}
\end{figure}

It is hard to predict what kind of semi-major axis and eccentricity distribution planet-planet
scattering would produce, but we assume that if the distribution of single-planet systems is
dominated by planet-planet scattering, and if the additional planets are ejected or are now beyond
$\sim 5$ au, the eccentricity of the close-in planets must be large.  We consider an initial
eccentricity distribution that peaks at $e \sim 0.6$ and that has a half-width of 0.1. We consider
3 different initial radial distributions, that shown in Figure \ref{a_initial}, and 2 from
Figure \ref{a_initial_peaked} (solid line and dotted line).  We integrate these
systems in the same way as before. The results are shown in Figure \ref{a_final_highe}. The initially unpeaked
distribution is shown as the solid-line.  The initially weakly peaked distribution is the dotted line, while the 
initially strongly peaked distribution is the dash-dot line. A larger 
fraction of the close-in planets are destroyed when compared with the simulations in which the initial
eccentricity peaked at $e = 0$ and so the initially weakly peaked distribution doesn't show a peak
after tidal evolution.  That    
the most peaked initial distribution (dash-dot line) has a pile-up that is similar to that
observed at least suggests that there could be an initial distribution that could result in a match with the
observed distribution.  However, tidal evolution also broadens the peak which suggests that the initial peak would need
to be narrower than we assume.

Although the above results do not preclude the possibility that the radial distribution of the observed 
single-planets systems which includes a peak at $\sim 3$ days is due primarily to planet-planet
scattering, it does suggest that this would require that planet-planet scattering is very effective
at scattering planets into orbits with $a < 0.1$ au which is not entirely consistent with theoretical
expectations \citep{ford01, marzari02}. Even \citet{nagasawa08}, who show that the Kozai mechanism
can significantly enhance the number of close-in planets, conclude that main channel for forming
close-in giant planets is probably still through Type II migration.  Furthermore, in these simulations,
the planets with $a > 0.1$ au undergo very little eccentricity evolution.  That the observed sample
of exoplanets has an eccentricity distribution that peaks at $e = 0$ \citep{wright09} suggests that the 
initial eccentricity distribution, at least for those with $a > 0.1$, must have also peaked at $e = 0$
which may suggest that the primary mechanism for getting these planets close to their parents stars
cannot be planet-planet scattering unless something else (other than tidal evolution) then reduces
their initial eccentricities.

\subsection{Mass-Period relation}
It has been suggested that the observed mass-period relation of close-in planets may give us some information
about their evolution.  \citet{ford06} suggest that if planets evolve from very high initial eccentricities, their
final orbital distance will be about twice the Roche distance (i.e., $a \sim 2 a_R$).  \citet{pont11} suggest - by 
studying the mass-period relation of transiting planets - that tidal circularization and the stopping mechanism
for close-in planets must be closely related. Our results here suggest that close-in planets are not typically
circularised from highly eccentric orbits and also that any stopping mechanism occurs prior to tidal evolution
playing a significant role. 

\begin{figure}
\begin{center}
\psfig{figure=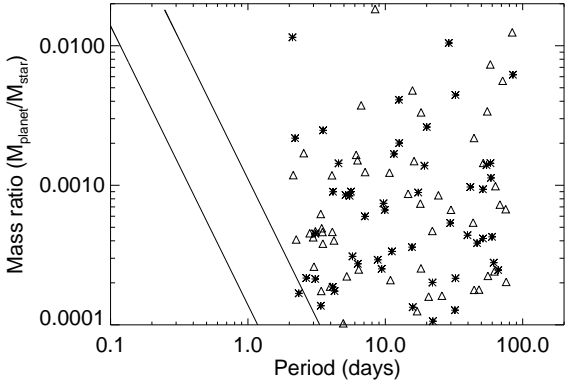,width=0.45\textwidth}
\end{center}
\caption{Scatter plot showing the mass-period relation for a randomly selected sample of
the modelled systems (asterisks) together with the mass-period relation for the observed systems (triangles).
The model we have chosen is the one in which there was an initial modest pile-up at $a \sim 0.05$ au (
solid line in Figure \ref{a_initial_peaked}). For the observed systems, we have only considered those detected
via the radial velocity technique.  The two diagonal lines are the Roche limit and twice the Roche limit.
The two populations appear, qualitatively, to be very similar.}
\label{scatter}
\end{figure}

The initial semi-major axis distribution that, after tidal circularisation, produced the best fit when compared
with the observed distribution of close-in planets was one in which there was a modest initial pile-up at $a \sim 0.05$ au 
(solid line in Figure \ref{a_initial_peaked}).  In Figure \ref{scatter} we plot the planet-to-star mass ratio against
orbital period for a randomly selected sample of these planets after they have undergone tidal evolution (asterisks) together with
the observed close-in planets detected via radial velocity (triangles). For the modelled systems we have randomly
selected the same number of planets as the number of observed close-in planets.  The two diagonal lines are
the Roche limit and twice the Roche limit.  Qualitatively, the mass-period relation of the modelled systems looks very similar 
to that of the observed systems.  Even though
our modelled sample was not tidally evolved from an initially highly eccentric orbit, the inner boundary is just beyond
twice the Roche limit. Admittedly, the sample is quite small and we have not analysed this in extensive details.  It does, however, 
appear as though evolution through Type II migration (with a mechanism for stopping planets at $a \sim 0.05$ au) followed by
tidal evoluion can produce a mass-period relation that is consistent with that observed. It has also been suggested \citep{davis09}
that close-in planets can also be lost through evaporation.  Although evaporation may well be operating, the lack of planets
close to their Roche limit is a natural consequence of inward migration followed by tidal evolution.

\section{Discussion and Conclusions}
The primary goal of this paper was to determine if the current distribution of close-in, gas-giant
($M > 0.1$ M$_{\rm Jup}$) exoplanets around Solar-like ($0.75$ M$_\odot \le$ M$_* \le 1.25$ M$_\odot$) 
stars is consistent with an initial distribution resulting
primarily from gap-opening migration in a gas disc (Type II) (e.g., \citealt{armitage02,armitage07}). 
The chosen initial distributions were evolved using tidal evolution equations \citep{dobbs04} for
a randomly chosen time between $2 \times 10^9$ and $6 \times 10^9$ years.  

The results suggests that an initial radial distribution due to Type II migration alone does not match the 
currently observed radial distribution which shows a pile-up of planets at $a \sim 0.05$ au.  If, however,
the inner regions of the disc are truncated due to interactions with the star's magnetosphere, it has been
suggested that planets should pile-up when inside the 2:1 resonance with the inner disc edge \citep{lin96,rice08} 
which would correspond to a semi-major axis of $\sim 0.05$ au for Solar-like stars with initial rotation periods of
$\sim 8$ days. The present study shows that the mass-period-eccentricity relation is indeed consistent with expectations
from disc migration with a magnetospheric gap and tides, at least for stars with masses in the
range between $0.75$ M$_\odot$ and $1.25$ M$_\odot$ and as long as the primordial pile-up is slightly (but not significantly) enhanced 
compared to what would be expected based on a steady-state model of Type II migration.  We also find a good, qualitative,
agreement between the mass-period relation of the modelled systems and that of the observed systems. 

This pile-up is, however, only evident for the lower-mass planets ($M_p < 2$ M$_{\rm Jup}$).  Simulated systems with a mass-independent
initial eccentricity distribution retain an initial pile-up for all planet masses.  The pile-up for higher-mass planets is, however, significantly
reduced if the initial eccentricity distribution is assumed to be mass dependent with the higher-mass planets preferentially having higher eccentricities.  
This is consistent with simulations \citep{rice08} suggesting that higher-mass planets will have enhanced eccentricity growth inside a
magnetospheric cavity.

We should stress that this does not prove that the primary mechanism for producing the observed close-in giant exoplanets 
is Type II migration but simply shows that - for reasonable assumptions about the initial distributions - it is consistent with 
this being the case.  We also consider one set of simulations with an initial eccentricity distribution that peaks at a large
eccentricity, aimed to mimic a possible planet-planet scattering scenario.  Although it can result in a final distribution
that matches that observed, it requires a more significant initial pile-up when compared to an initial eccentricity
distribution that peaks at $e = 0$.  Furthermore, the planets beyond $a \sim 0.1$ au retain their high eccentricities
which is not consistent with observations.  

The observation of misaligned, and in some cases retrograde, close-in planets \citep{winn09,colliercameron10, triaud10} does, however, strongly suggest
that planet-planet scattering (which may undergo ``Kozai" cycles if a binary or distant planetary companion is present) must
play a role in the formation of some close-in exoplanets.  It is intriguing that misaligned systems occur at all masses,
but dominate for stars with $T_{\rm eff} > 6250$ K \citep{winn10,barnes11}.  In their population-synthesis studies that include
disc migrations \citet{alibert11} find that the short disc lifetimes of high-mass stars doesn't leave enough time for Jupiter-mass
planets to migrate into close orbits.  Our results, therefore, appear to be consistent with this.  For stars with masses between $0.75$ and
$1.25$ M$_\odot$, the distribution of close-in giant planets can be modelled by assuming Type II migration followed by tidal
evolution.  There will be some systems that have been influenced by planet-planet scattering, which may have undergone ``Kozai" cycles, but these
do not dominate.  For the more massive stars ($M_* > 1.25$ M$_\odot$), where the disc lifetime is too short for Type II migration
to be effective, close-in giant planets are preferentially formed through dynamical interactions in which ``Kozai" cycles may also
operate.  

\section*{acknowledgements}
The authors would like to thank the anonymous referee for helpful comments that have improved this paper.
W.K.M.R. acknowledges support from the Scottish Universities Physics Alliance (SUPA) 
and for support from the Science and Technology Facilities Council (STFC) through
grant ST/H002380/1. The authors would also like to thank the Isaac Newton Institute for Mathematical
Sciences for their hospitality during the Dynamics of Discs and Planets Programme and would
like to thank Phil Armitage for many useful discussions.

\end{document}